\documentclass[english]{article}
\usepackage{times}
\usepackage{url}
\usepackage{bbm}
\usepackage[T1]{fontenc}
\usepackage[latin1]{inputenc}
\usepackage{geometry}
\usepackage{rotating}
\usepackage{color}
\usepackage{graphicx}
\usepackage{amsmath, amsthm, amssymb}
\usepackage{setspace}
\usepackage{lineno}
\usepackage{hyperref}
\usepackage{bbm}
\usepackage{natbib} \bibpunct{(}{)}{;}{author-year}{}{,} 

\usepackage{color}
\definecolor{trustcolor}{rgb}{0,0,1}

\makeatletter

\usepackage{babel}
\makeatother
\begin{document}

\title{Detecting genomic signatures of natural selection with principal component analysis: application to the 1000 Genomes data}

\author{Nicolas Duforet-Frebourg$^{1,2,3}$, Keurcien Luu$^{1,2}$, Guillaume Laval$^{4,5}$, Eric Bazin$^{6}$, Michael G.B. Blum$^{1,2,*}$}

\date{~ }

\maketitle
\noindent$^{1}$ Univ. Grenoble Alpes, TIMC-IMAG UMR 5525, F-38000 Grenoble, France\\
\noindent$^{2}$ CNRS, TIMC-IMAG, F-38000 Grenoble, France\\
\noindent$^{3}$ Department of Integrative Biology, University of California, Berkeley, California 94720-3140, USA\\
\noindent$^{4}$ Institut Pasteur, Human Evolutionary Genetics, Department of Genomes and Genetics, Paris, France\\
 \noindent$^{5}$ Centre National de la Recherche Scientifique, URA3012, Paris, France\\
\noindent$^{6}$ Univ. Grenoble Alpes, CNRS, Laboratoire d'Ecologie Alpine UMR 5553, F-38000 Grenoble, France


\noindent Running Head:  Detection of positive selection based on principal component analysis

Corresponding author:
Michael  Blum

Laboratoire TIMC-IMAG, Facult\'e de M\'edecine, 38706 La Tronche, France 

Phone +33 4 56 52 00 65

Email: michael.blum@imag.fr

\clearpage

\begin{abstract}
{\normalsize
To characterize natural selection, various analytical methods for detecting candidate genomic regions have been developed. We propose to perform genome-wide scans of natural selection using principal component analysis. We show that the common $F_{ST}$ index of genetic differentiation between populations can be viewed as the proportion of variance explained by the principal components. Considering the correlations between genetic variants and each principal component provides a conceptual framework to detect genetic variants involved in local adaptation without any prior definition of populations. To validate the PCA-based approach, we consider the 1000 Genomes data (phase 1) considering 850 individuals coming from Africa, Asia, and Europe. The number of genetic variants is of the order of $36$ millions obtained with a low-coverage sequencing depth (3X). The correlations between genetic variation and each principal component provide well-known targets for positive selection (EDAR, SLC24A5, SLC45A2, DARC), and also new candidate genes (APPBPP2, TP1A1, RTTN, KCNMA, MYO5C) and non-coding RNAs. In addition to identifying genes involved in biological adaptation, we identify two biological pathways involved in polygenic adaptation that are related to the innate immune system (beta defensins) and to lipid metabolism (fatty acid omega oxidation). An additional analysis of European data shows that a genome scan based on PCA retrieves classical examples of local adaptation even when there are no well-defined populations. PCA-based statistics, implemented in the {\it PCAdapt} \verb R  package and the {\it PCAdapt} open-source software, retrieve well-known signals of human adaptation, which is encouraging for future whole-genome sequencing project, especially when defining populations is difficult. 
}
\end{abstract}
\clearpage

\section*{Significance statement}
Positive natural selection or local adaptation is the driving force behind the adaption of individuals to their environment. To identify genomic regions responsible for local adaptation, we propose to consider the genetic markers that are the most related with population structure. To uncover genetic structure, we consider principal component analysis that identifies the primary axes of variation in the data. Our approach generalizes common approaches for genome scan based on measures of population differentiation. To validate our approach, we consider the human 1000 Genomes data and find well-known targets for positive selection as well as new candidate regions. We also find evidence of polygenic adaptation for two biological pathways related to the innate immune system and to lipid metabolism. 


\section*{Introduction}
Because of the flood of genomic data, the ability to understand the genetic architecture of natural selection has dramatically increased. Of particular interest is the study of local positive selection which explains why individuals are adapted to their local environment. In humans, the availability of genomic data fostered the identification of loci involved in positive selection \cite[]{sabeti07,barreiro08,pickrell09,grossman13}. Local positive selection tends to increase genetic differentiation, which can be measured by difference of allele frequencies between populations \cite[]{sabeti06,nielsen05,colonna14}. For instance, a mutation in the DARC gene that confers resistance to malaria is fixed in Sub-Saharan African populations whereas it is absent elsewhere \cite[]{hamblin02}. In addition to the variants that confer resistance to pathogens, genome scans also identify other genetic variants, and many of these are involved in human metabolic phenotypes and morphological traits \cite[]{barreiro08,hancock10}.

In order to provide a list of variants potentially involved in natural selection, genome scans compute measures of genetic differentiation between populations and consider that extreme values correspond to candidate regions \cite[]{luikart03}. The most widely used index of genetic differentiation is the $F_{ST}$ index which measures the amount of genetic variation that is explained by variation between populations \cite[]{excoffier92}. However the $F_{ST}$ statistic requires to group individuals into populations which can be problematic when ascertainment of population structure does not  show well-separated clusters of individuals \cite[e.g.][]{novembre08b}. Other statistics related to $F_{ST}$ have been derived to reduce the false discovery rate obtained with $F_{ST}$ but they also work at the scale of populations  \cite[]{bonhomme10,fariello13,gunther13}. Grouping individuals into populations can be subjective, and important signals of selection may be missed with an inadequate choice of populations \cite[]{yang12}. We have previously developed an individual-based approach for selection scan based on a Bayesian factor model but the MCMC algorithm required for model fitting does not scale well to large data sets containing a million of variants or more \cite[]{duforet14}.

We propose to detect candidates for natural selection using principal component analysis (PCA). PCA is a technique of multivariate analysis used to ascertain population structure \cite[]{patterson06}. PCA decomposes the total genetic variation into $K$ axes of genetic variation called principal components. In population genomics, the principal components can correspond to evolutionary processes such as evolutionary divergence between  populations  \cite[]{mcvean09}. Using simulations of an island model and of a model of population fission followed by isolation, we show that the common $F_{ST}$ statistic corresponds to the proportion of variation explained by the first $K$ principal components when $K$ has been properly chosen. With this point of view, the $F_{ST}$ of a given variant is obtained by summing the squared correlations of the first $K$ principal components opening the door to new statistics for genome scans. At a genome-wide level, it is known that there is a relationship between $F_{ST}$ and PCA \cite[]{mcvean09}, and our simulations show that the relationship also applies at the level of a single variant. 

The advantages of performing a genome scan based on PCA are multiple: it does not require to group individuals into populations, the computational burden is considerably reduced compared to genome scan approaches based on MCMC algorithms  \cite[]{foll08,riebler08,gunther13,duforet14}, and candidate SNPs can be related to different evolutionary events that correspond to the different PCs. Using simulations and the 1000 Genomes data, we show that PCA can provide useful insights for genome scans. Looking at the correlations between SNPs and principal components provides a novel conceptual framework to detect genomic regions that are candidates for local adaptation.

\section*{New method}
\subsection*{New statistics for genome scan}

We denote by ${\bf Y}$ the $(n\times p)$ centered and scaled genotype matrix where $n$ is the number of individuals and $p$ is the number of loci. The new statistics for genome scan are based on principal component analysis. The objective of PCA is to find a new set of orthogonal variables called the principal components,  which are linear combinations of (centered and standardized) allele counts, such that the projections of the data onto these axes lead to an optimal summary of the data. To present the method, we introduce the truncated singular value decomposition (SVD) that approximates the data matrix ${\bf Y}$ by a matrix of smaller rank
\begin{equation}
\label{eq:svd}
{\bf Y}\approx{\bf U}{\bf \Sigma} {\bf V}^T,
\end{equation}
where ${\bf U}$ is a $(n\times K)$ orthonormal matrix, ${\bf V}$ is a $(p\times K)$ orthonormal matrix, ${\bf \Sigma}$ is a diagonal $(K\times K)$ matrix and $K$ corresponds to the rank of the approximation. 
The  solution of PCA with $K$ components can be obtained using the truncated SVD of equation (\ref{eq:svd}) : the $K$ columns of ${\bf V}$ contain the coefficients of the new orthogonal variables, the $K$ columns of ${\bf U}$ contain the projections (called {\it scores}) of the original variables onto the principal components and capture population structure (Fig. S1), and the squares of the elements of ${\bf \Sigma}$ are proportional to the  proportion of variance explained by each principal component \cite[]{jolliffe05}. We denote the diagonal elements of ${\bf \Sigma}$ by $\sqrt \lambda_k$, $k=1,\dots,K$ where the $\lambda_k$'s are the ranked eigenvalues of the matrix ${\bf Y} {\bf Y}^{T}$. Denoting by ${\bf V}_{jk}$, the entry of ${\bf V}$ at the $j^{th}$ line and $k^{th}$ column, then the correlation $\rho_{jk}$ between the $j^{\rm th}$ SNP and the $k^{\rm th}$ principal component is given by $\rho_{jk}= \sqrt \lambda_k V_{jk}/\sqrt{n-1}$ \cite[]{cadima95}. In the following, the statistics $\rho_{jk}$ are referred to as {\it loadings} and will be used for detecting selection.

The second statistic we consider for genome scan corresponds to the proportion of variance of a SNP that is explained by the first $K$ PCs. It is called the communality in exploratory factor analysis because it is the variance of observed variables accounted for by the common factors, which correspond to the first $K$ PCs \cite[]{suhr2009}.  Because the principal components are orthogonal to each other, the proportion of variance explained by the first $K$ principal components is equal to the sum of the squared correlations with the first $K$ principal components. Denoting by $h_j^2$ the communality of the $j^{th}$ SNP, we have
\begin{equation}
\label{eq:h}
h_j^2=\sum_{k=1}^K \rho_{jk}^2.
\end{equation}

The last statistic we consider for genome scans sums the squared of normalized loadings. It is defined as $h_j^{\prime2}=\sum_{k=1}^K V_{jk}^2$. Compared to the communality $h^2$, the statistic $h^{\prime2}$ should theoretically give the same importance to each PC because the normalized loadings are on the same scale as we have $\sum_{j=1}^p V_{jk}^2 =1$, for $k=1\dots K$.  

\subsection*{Numerical computations}
The method of selection scan should be able to handle a large number $p$ of genetic variants. In order to compute truncated SVD with large values of $p$, we compute the $n\times n$ covariance matrix ${\bf \Omega}={{\bf YY}^T}/(p-1)$. The covariance matrix ${\bf \Omega}$ is typically of much smaller dimension than the $p\times p$ covariance matrix. Considering the  $n\times n$ covariance matrix ${\bf \Omega}$ speeds up matrix operations. Computation of the covariance matrix is the most costly operation and it requires a number  of arithmetic operations proportional to $p n^2$. After computing the covariance matrix ${\bf \Omega}$,  we compute its first $K$ eigenvalues and eigenvectors to find $\Sigma^2/(p-1)$ and ${\bf U}$. Eigenanalysis is performed with the {\it dsyevr} routine of the linear algebra package LAPACK \cite[]{lapack}. The matrix ${\bf V}$, which captures the relationship between each SNPs and population structure, is obtained by the matrix operation ${\bf V}^T= {\bf \Sigma}^{-1}{\bf U}^T {\bf Y}$ that arises from equation (\ref{eq:svd}). In the software {\it PCAdapt}, data are processed as a stream and never stored in order to have a very low  memory  access whatever the size of the data.

\section*{Results}

\subsection*{Island model}

To investigate the relationship between communality $h^2$ and $F_{ST}$, we consider an island model with three islands. We use $K=2$ when performing PCA because there are 3 islands. We choose a value of the migration rate that generates a mean $F_{ST}$ value (across the $1,400$ neutral SNPs) of $4\%$.  We consider five different simulations with varying strengths of selection for the $100$ adaptive SNPs. In all simulations, the $R^2$ correlation coefficient between $h^2$ and $F_{ST}$ is larger than $98\%$. Considering as candidate SNPs the one percent of the SNPs with largest values of $F_{ST}$ or of $h^2$, we find that the overlap coefficient between the two sets of SNPs is comprised between $88\%$ and $99\%$. When varying the strength of selection for adaptive SNPs, we find that the relative difference of false discovery rates (FDR) obtained with $F_{ST}$ (top $1\%$) and with $h^2$ (top $1\%$) is smaller than $5\%$. The similar values of FDR obtained with $h^2$ and with $F_{ST}$ decrease for increasing strength of selection (Fig. S2).

\subsection*{Divergence model}

\begin{figure}[ht!]
	\centering
		\includegraphics[width=14cm]{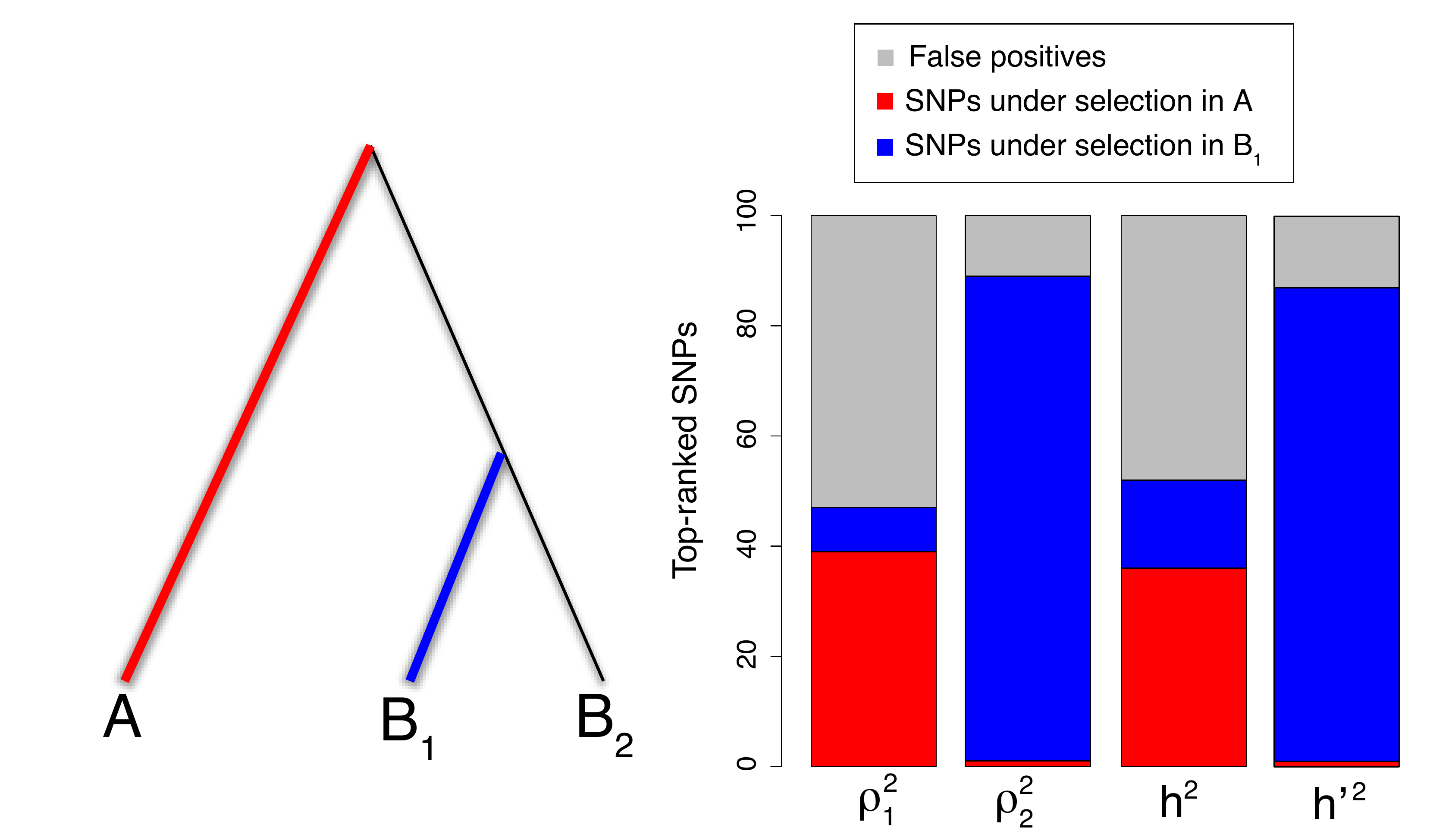}
	\caption{Repartition of the $1\%$ top-ranked SNPs for each PCA-based statistic under a divergence model with two types of adaptive constraints. Thicker and colored lineages correspond to lineages where adaptation took place. The squared loadings with PC1 $\rho_{j1}^2$ pick a large proportion of SNPs involved in selection in population $A$ whereas the squared loadings with PC2 $\rho_{j2}^2$ pick SNPs involved in selection in population $B_1$. This difference is reflected in the different repartition of the top-ranked SNPs for the communality $h^2$ and the statistic $h^{\prime2}$.}
 \label{fig:stacked_bar}
\end{figure}

\begin{figure}[ht!]
	\centering
		\includegraphics[width=14cm]{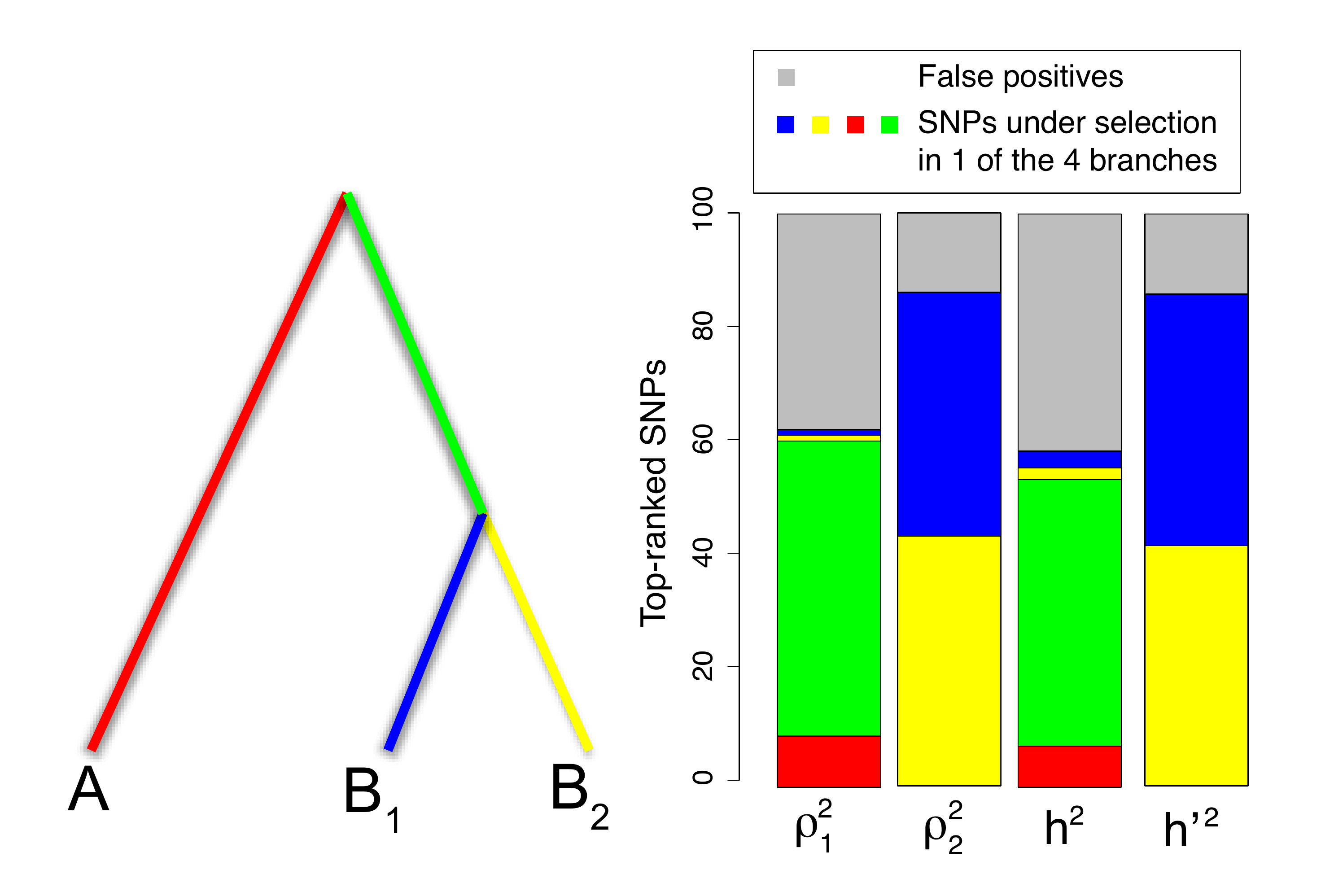}
	\caption{Repartition of the $1\%$ top-ranked SNPs of each PCA-based statistic under a divergence model with four types of adaptive constraints. Thicker and colored lineages correspond to lineages where adaptation occurred. The different types of SNPs picked by the squared loadings $\rho_{j1}^2$ and $\rho_{j2}^2$ is also found when comparing the communality $h^2$ and the statistic $h^{\prime 2}$.}
	 \label{fig:stacked_bar2}
\end{figure}

To compare the performance of different PCA-based summary statistics, we simulate genetic variation in models of population divergence. The divergence models assume that there are three populations, $A$, $B_1$ and $B_2$ with $B_1$ and $B_2$ being the most related populations (Figs. \ref{fig:stacked_bar} and \ref{fig:stacked_bar2}). The first simulation scheme assumes that local adaptation took place in the lineages corresponding to the environments  of populations $A$ and $B_1$ (Fig. \ref{fig:stacked_bar}). The SNPs, which are are assumed to be independent, are divided into 3 groups: 9,500 SNPs evolve neutrally, 250 SNPs confer a selective advantage in the environment of $A$, and 250 other SNPs confer a selective advantage in the environment of $B_1$. Genetic differentiation, measured by pairwise $F_{ST}$, is equal to $14\%$ when comparing population $A$ to the other ones and is equal to $5\%$ when comparing populations $B_1$ and $B_2$. Performing principal component analysis with $K=2$ shows that the first component separates 
population $A$ from $B_1$ and $B_2$ whereas the second component separates $B_1$ from $B_2$ (Fig. S1). The choice of $K=2$ is evident when looking at the scree plot because the eigenvalues, which are proportional to the proportion of variance explained by each PC, drop beyond $K=2$ and stay almost constant as $K$ further increases (Fig. S3).

We investigate the relationship between the communality statistic $h^2$, which measures the proportion of variance explained by the first two PCs, and the $F_{ST}$ statistic. We find a squared Pearson correlation coefficient between the two statistics larger than $98.8\%$  in the simulations corresponding to Figs. \ref{fig:stacked_bar} and \ref{fig:stacked_bar2} (Fig. S4). For these two simulations, we look at the  SNPs in the top $1\%$ (respectively $5\%$) of the ranked lists based on $h^2$ and $F_{ST}$, and we find an overlap coefficient always larger than $93\%$ for the lists provided by the two different statistics (respectively $95\%$).  Providing a ranking of the SNPs almost similar to the ranking provided by $F_{ST}$ is therefore possible without considering that individuals originate from predefined populations.

We then compare the performance of the different statistics based on PCA by investigating if the top-ranked SNPs (top $1\%$) manage to pick SNPs involved in local adaptation (Fig. \ref{fig:stacked_bar}). The squared loadings $\rho^2_{j1}$ with the first PC pick SNPs involved in selection in population $A$ ($39\%$ of the top $1\%$), a few SNPs involved  in selection in $B_1$ ($9\%$), and many false positive SNPs (FDR of $53\%$). The squared loadings with the second PC $\rho^2_{j2}$ pick less false positives (FDR of $12\%$) and most SNPs are involved in selection in $B_1$ ($88\%$) with just a few involved in selection in $A$ ($1\%$). When adaptation took place in two different evolutionary lineages of a divergence tree between populations, a genome scan based on PCA has the nice property that outlier loci correlated with PC1 or with PC2 correspond to adaptive constraints that occurred in different parts of the tree.

Because the communality $h^2$ gives more importance to the first PC (equation(\ref{eq:h})), it picks preferentially the SNPs that are the most correlated with PC1. There is a large overlap of $72\%$ between the $1\%$ top-ranked lists provided by $h^2$ and $\rho^2_{j1}$. Therefore, the communality statistic $h^2$ is more sensitive to ancient adaptation events that occurred in the  environment of population $A$. By contrast, the alternative statistic $h^{\prime2}$ is more sensitive to recent adaptation events that occurred in the environment of population $B_1$. When considering the top-ranked $1\%$ of the SNPs, $h^{\prime2}$ captures only one SNP involved in selection in $A$ ($1\%$ of the top $1\%$) and 88 SNPs related to adaptation in $B_1$ ($88\%$ of the top $1\%$). The overlap between the $1\%$ top-ranked lists provided by $h'^2$ and by $\rho^2_{j2}$ is of $86\%$.

The $h^{\prime2}$ statistic is mostly influenced by the second principal component because the distribution of squared loadings corresponding to the second PC has a heavier tail, and this result holds for the two divergence models and for the 1000 Genomes data (Fig. S5). To summarize, the $h^2$ and $h^{\prime2}$ statistics give too much importance to PC1 and PC2 respectively and they fail to capture in an equal manner both types of adaptive events occurring in the environment of populations $A$ and $B_1$.

We also investigate a more complex simulation in which adaptation occurs in the four branches of the divergence tree (Fig. \ref{fig:stacked_bar2}). Among the $10,000$ simulated SNPs, we assume that there are four sets of $125$ adaptive SNPs with each set being related to adaptation in one of the four branches of the divergence tree. Compared to the simulation of Fig. \ref{fig:stacked_bar}, we find the same pattern of population structure (Fig. S1). The squared loadings $\rho^2_{j1}$ with the first PC mostly pick SNPs involved in selection in the branch that predates the split between $B_1$ and $B_2$ ($51\%$ of the top $1\%$), SNPs involved in selection in the environment of population $A$ ($9\%$), and false positive SNPs (FDR of $38\%$). Except for false positives (FDR of $14\%$),  the squared loadings $\rho^2_{j2}$ with the second PC rather pick SNPs involved in selection in $B_1$ and $B_2$ ($42\%$ for $B_1$ and $44\%$ for $B_2$). Once again, there is a large overlap between the SNPs picked by the communality $h^2$ and by $\rho^2_{1}$ ($92\%$ of overlap) and between the SNPs picked by $h'^2$ and $\rho^2_{2}$ ($93\%$ of overlap). Because the first PC discriminates population $A$ from $B_1$ and $B_2$ (Fig. S1), the SNPs most correlated with PC1 correspond to SNPs related to adaptation in the (red and green) branches that separate $A$ from populations $B_1$ and $B_2$. By contrast, the SNPs that are most correlated to PC2 correspond to SNPs related to adaptation in the two (blue and yellow) branches that separate population $B_1$ from $B_2$ (Fig. \ref{fig:stacked_bar2}).

We additionally evaluate to what extent the results are robust with respect to some parameter settings.  When considering the $5\%$ of the SNPs with most extreme values of the statistics instead of the top $1\%$, we also find that the summary statistics pick SNPs related to different evolutionary events (Fig. S6). The main difference being that the FDR increases considerably when considering the top $5\%$ instead of the top $1\%$(Fig. S6). We also consider variation of the selection coefficient ranging from $s=1.01$ to $s=1.1$ ($s=1.025$ corresponds to the simulations of Figures \ref{fig:stacked_bar} and \ref{fig:stacked_bar2}). As expected, the false discovery rate of the different statistics based on PCA is considerably reduced when the selection coefficient increases (Fig. S7).

In the divergence model of Fig \ref{fig:stacked_bar}, we also compare the false discovery rates obtained with the statistics $h^2$, $h^{\prime2}$ and with a Bayesian  factor model implemented in the software {\it PCAdapt} \cite[]{duforet14}. For the optimal choice of $K=2$, the statistic $h^{\prime2}$ and the Bayesian factor model provide the smallest FDR (Fig. S8). However, when varying the value of $K$ from $K=1$ to $K=6$, we find that the communality $h^2$ and the Bayesian approach are robust to over-specification of $K$ ($K>3$)  whereas the false discovery rate obtained with $h^{\prime2}$ increases importantly as $K$ increases beyond $K=2$ (Fig. S8).

We also consider a more general isolation-with-migration model. In the divergence model where adaptation occurs in two different lineages of the population tree (Figure \ref{fig:stacked_bar}), we add constant migration between all pairs of populations. We assume that migration occurred after the split between $B_1$ and $B_2$. We consider different values of migration rates generating a mean $F_{ST}$ of $7.5\%$ for the smallest migration rate to a mean $F_{ST}$ of $0\%$ for the largest migration rate. We find that the $R^2$ correlation between $F_{ST}$ and $h^2$ decreases as a function of the migration rate (Fig S9). For $F_{ST}$ values larger than $0.5\%$, $R^2$ is larger than $97\%$. The squared correlation $R^2$ decreases to $47\%$ for the largest migration rate. Beyond a certain level of migration rate, population structure, as ascertained by principal components, is no more described by well-separated clusters of individuals (Fig S10) but by a more clinal or continuous pattern (Fig S10) explaining the difference between $F_{ST}$ and $h^2$.  However, the false discovery rates obtained with the different statistics based on PCA and with $F_{ST}$ evolve similarly as a function of the migration rate. For both types of approaches, the false discovery rate increases for larger migration with almost no true discovery (only 1 true discovery in the top $1\%$ lists) when considering the largest migration rate.

The main results obtained under the divergence models can be described as follows. The principal components correspond to different evolutionary lineages of the divergence tree. The communality statistic $h^2$ provides similar list of candidate SNPs than $F_{ST}$ and it is mostly  influenced by the first principal component which can be problematic if other PCs also convey adaptive events. To counteract this limitation, which can potentially lead to the loss of important signals of selection, we show that looking at the squared loadings with each of the principal components provide adaptive SNPs that are related to different evolutionary events. When adding migration rates between lineages, we find that the main results are unchanged up to a certain level of migration rate. Above this level of migration rate, the relationship between $F_{ST}$  and $h^2$ does not hold anymore and  genome scans based on either PCA or $F_{ST}$ produce a majority of false positives.

\subsection*{1,000 Genome data}
\begin{figure}[ht!]
	\centering
		\includegraphics[width=14cm]{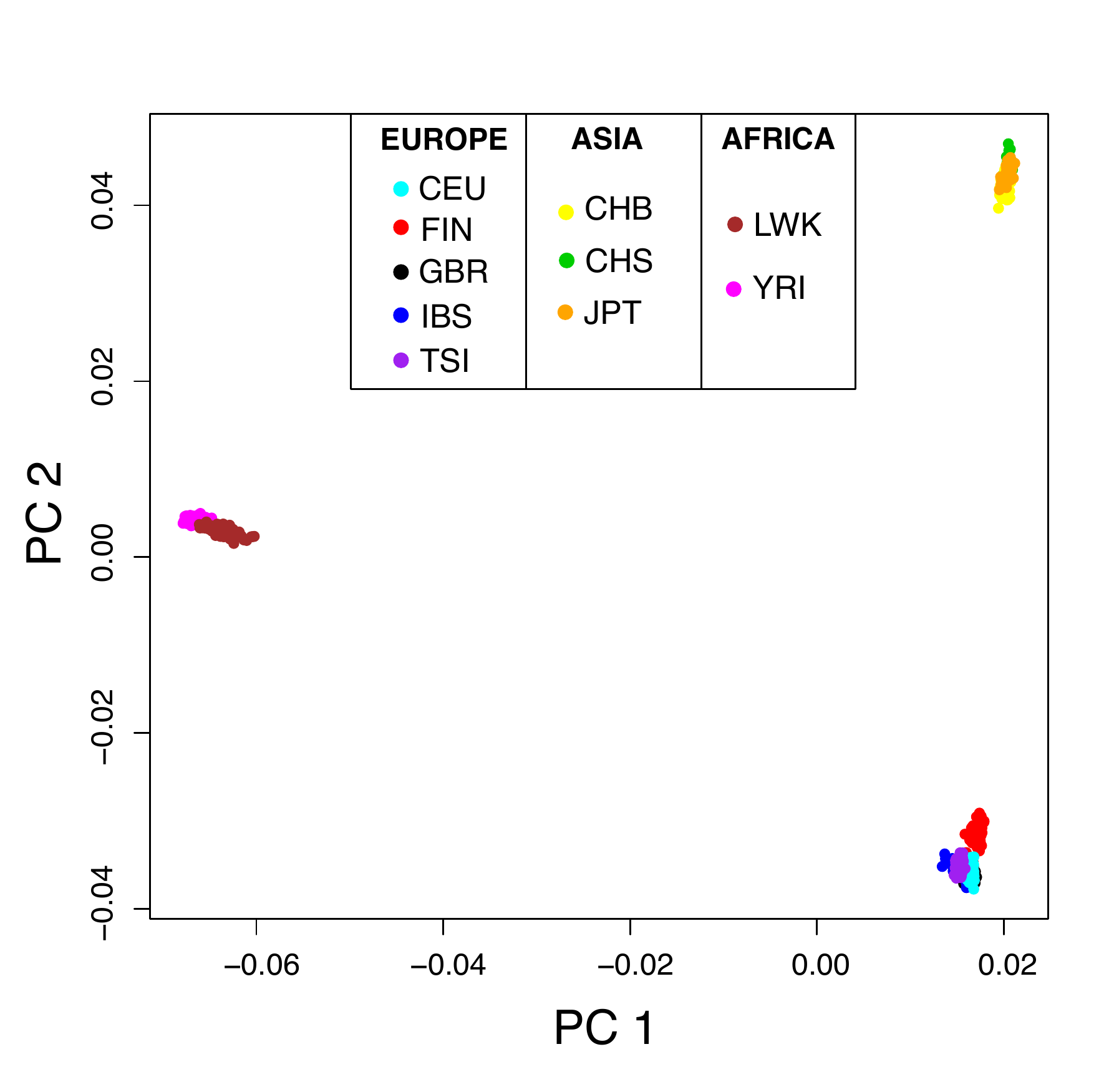}
	\caption{Principal component analysis with $K=2$ applied to the 1000 Genomes data. The sampled populations are the following: British in England and Scotland (GBR), Utah residents with Northern and Western European ancestry (CEU), Finnish in Finland (FIN), Iberian populations in Spain (IBS), Toscani in Italy (TSI), Han Chinese in Bejing (CHB), Southern Han Chinese (CHS), Japanese in Tokyo (JPT), Luhya in Kenya (LWK), Yoruba in Nigeria (YRI).}
 \label{fig:PC}
\end{figure}

Since we are interested in selective pressures that occurred during the human diaspora out of Africa, we decide to exclude individuals whose genetic makeup is the result of recent admixture events (African Americans, Columbians, Puerto Ricans and Mexicans). The first three principal  components capture population structure whereas the following components separate individuals within populations (Figs. \ref{fig:PC} and S11). The first and second PCs ascertain population structure between Africa, Asia and Europe (Fig. \ref{fig:PC}) and the third principal component separates the Yoruba from the Luhya population (Fig. S11). The decay of eigenvalues suggests to use $K=2$ because the eigenvalues drop between $K=2$ and $K=3$ where a plateau of eigenvalues is reached (Fig. S3).

When performing a genome scan with PCA, there are different choices of statistics. The first choice is the $h^2$ communality statistic. Using the three continents as labels, there is a squared correlation between $h^2$ and $F_{ST}$ of $R^2=0.989$. To investigate if $h^2$ is mostly influenced by the first PC, we determine if the outliers for the $h^2$ statistics are related with PC1 or with PC2. Among the  top $0.1\%$ of SNPs with the  largest values of $h^2$, we find that $74\%$ are in the top $0.1\%$ of the squared loadings $\rho_{j1}^2$ corresponding to PC1 and $20\%$ are in the top $0.1\%$ of the squared loadings $\rho_{j2}^2$ corresponding to PC2. The second possible choice of summary statistics is the $h^{\prime2}$ statistic. Investigating the repartition of the $0.1\%$ outliers for $h'$, we find that  $0.005\%$ are in the top $0.1\%$ of the squared loadings $\rho_{j1}^2$ corresponding to PC1 and $85\%$ are in the top $0.1\%$ of the squared loadings $\rho_{j2}^2$ corresponding to PC2. The $h^{\prime2}$ statistic is mostly influenced by the second PC because the distribution of the $V_{2j}^2$ (normalized squared loadings) has a longer tail than the corresponding distribution for PC1 (Fig. S5). Because the $h^2$ statistic is mostly influenced by PC1 and $h^{\prime2}$ is mostly influenced by PC2, confirming the results obtained under the divergence models, we rather decide to perform two separate genome scans based on the squared loadings $\rho_{j1}^2$ and $\rho_{j2}^2$.

\begin{figure}[ht!]
	\centering
		\includegraphics[width=14cm]{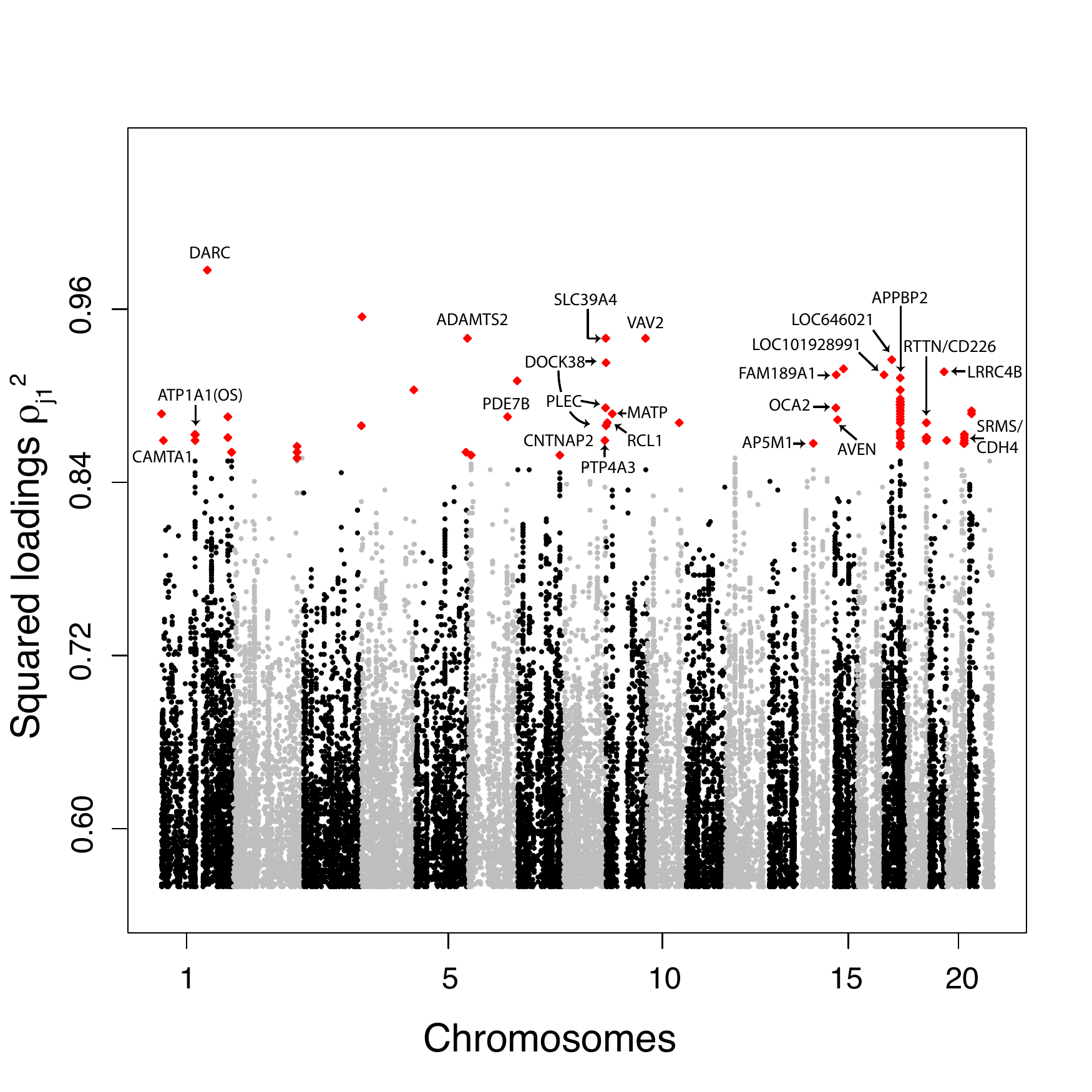}
	\caption{Manhattan plot for the 1000 Genomes data of the squared loadings $\rho_{j1}^2$ with the first principal component. For sake of presentation, only the top-ranked SNPs (top $0.1\%$) are displayed and the 100 top-ranked SNPs are colored in red.}
\label{fig:scan_PC1}.
\end{figure}

\begin{figure}[ht!]
	\centering
		\includegraphics[width=14cm]{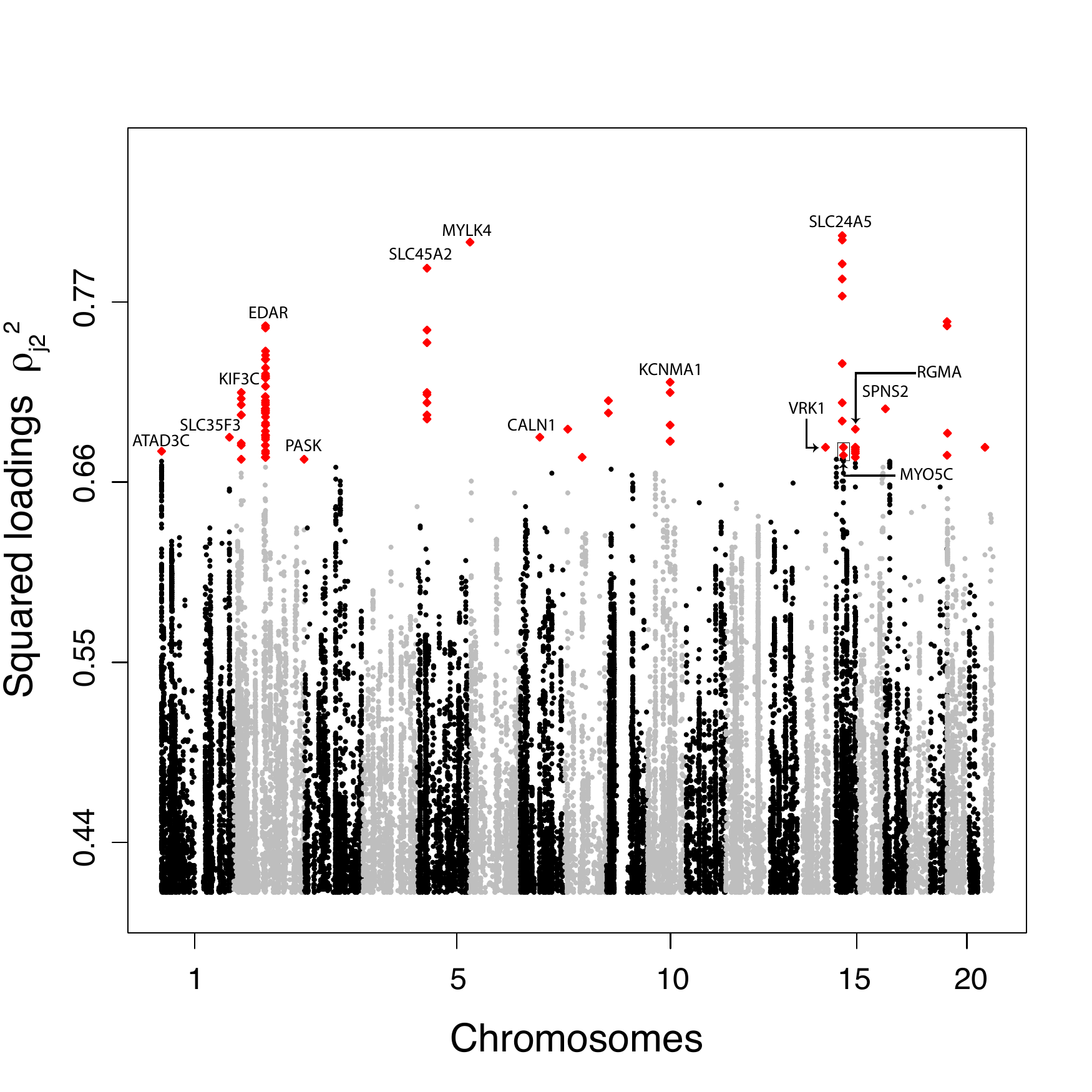}
	\caption{Manhattan plot for the 1000 Genomes data of the squared loadings $\rho_{j2}^2$ with the second principal component.  For sake of presentation, only the top-ranked SNPs (top $0.1\%$) are displayed and the 100 top-ranked SNPs are colored in red.}
\label{fig:scan_PC2}.
\end{figure}

The two Manhattan plots based on the squared loadings for PC1 and  PC2 are displayed in Figs. \ref{fig:scan_PC1} and \ref{fig:scan_PC2} (Table S1 contains the loadings for all variants). Because of Linkage Disequilibrium, Manhattan plots generally produce clustered outliers. To investigate if the top $0.1\%$ outliers are clustered in the genome, we count---for various window sizes---the proportion of contiguous windows containing at least one outlier. We find that outlier SNPs correlated with PC1 or with PC2 are more clustered than expected if they would have been uniformly distributed among the $36,536,154$ variants (Fig. S12). Additionally, the clustering is larger for the outliers related to the second PC as they cluster in fewer windows (Fig. S12). As the genome scan for PC2 captures more recent adaptive events, it reveals larger genomic windows that experienced fewer recombination events.

The 1,000 Genome data contain many low-frequency SNPs; $82\%$ of the SNPs have a minor allele frequency smaller than $5\%$. However, these low-frequency variants are not found among outlier SNPs. There are no SNP with a minor allele frequency smaller than $5\%$ among the $0.1\%$ of the SNPs most correlated with PC1 or with PC2.

The 100 SNPs that are the most correlated with  the first PC are located in 24 genomic regions (Table S2). Most of the regions contain just one or a few SNPs except a peak in the gene APPBP2 that contains 33 out of the 100 top SNPs, a peak encompassing the RTTN and CD226 genes containing 17 SNPS and a peak in the ATP1A1 gene containing 7 SNPs  (Fig. \ref{fig:scan_PC1}). Confirming a larger clustering for PC2 outliers, the 100 SNPs that are the most correlated with PC2 cluster in fewer genomic regions (Table S3). They are located in 14 genomic regions including a region overlapping with EDAR contains 44 top hits, two regions containing 8 SNPs and located in the pigmentation genes SLC24A5 and SLC45A2, and two regions with 7 top hit SNPs, one in the gene KCNMA1 and another one encompassing the RGLA/MYO5C genes (Fig.  \ref{fig:scan_PC2}).

We perform Gene Ontology enrichment analyses using {\it Gowinda}  for the SNPs that are the most correlated with PC1 and PC2. For PC1, we find, among others, enrichment (${\rm FDR} \leq 5\%$) for ontologies related to the regulation of  arterial blood pressure, the endocrine system and the immunity response (interleukin production, response to viruses) (Table S4). For PC2, we find enrichment ($FDR \leq 5\%$)  related to olfactory receptors, keratinocyte and epidermal cell differentiation, and ethanol metabolism (Table S5). We also search for polygenic adaptation by looking for biological pathways enriched with outlier genes \cite[]{daub13}. For PC1, we find one enriched (${\rm FDR} \leq 5\%$) pathway consisting of the beta defensin pathway (Table S6). The  beta defensin pathway contains mainly genes involved in the innate immune system consisting of 36 defensin genes and of 2 Toll-Like receptors (TLR1 and TLR2). There are additionally 2 chemokine receptors (CCR2 and CCR6) involved in the beta defensin pathway. For PC2, we also find one enriched pathway consisting of fatty acid omega oxidation (${\rm FDR} \leq 5\%$, Table S7). This pathway consists of genes involved in alcohol oxidation (CYP, ALD and ALDH genes). Performing a less stringent enrichment analysis which can find pathways containing overlapping genes, we find more enriched pathways: the beta defensin and the defensin pathways for PC1 and ethanol oxidation, glycolysis/gluconeogenesis and fatty acid omega oxidation for PC2 (Table S8).

To further validate the proposed list of candidate SNPs involved in local adaptation, we test for an enrichment of  genic or non-synonymous SNP among the SNPs that are the most correlated with the PC. We measure the enrichment  among outliers by computing odds ratio \cite[]{kudaravalli09,fagny14}. For PC1, we do not find significant enrichments (Table \ref{tab:1}) except when measuring the enrichment of genic regions compared to non-genic regions ($OR=10.18$ for the 100 most correlated SNPs, $P<5\%$ using a permutation procedure). For PC2, we find an enrichment of genic regions among outliers as well as an enrichment of non-synonymous SNPs (Table \ref{tab:1}). By contrast with the enrichment of genic regions for SNPs extremely correlated with the first PC, the enrichment for the variants extremely correlated with PC2 outliers is significant when using different thresholds to define outliers (Table \ref{tab:1}).

\section*{Discussion}
The promise of a fine characterization of natural selection in humans fostered the development of new analytical methods for detecting candidate genomic regions \cite[]{vitti13}. Population-differentiation based methods such as genome scans based on $F_{ST}$ look for marked differences in allele frequencies between population \cite[]{holsinger09}. Here, we show that the communality statistic $h^2$, which measures the proportion of variance of a SNP that is explained by the first $K$ principal components, provides a similar list of outliers than the $F_{ST}$ statistic when there are $K+1$ clusters of populations. In addition, the communality statistic $h^2$ based on PCA can be viewed as an extension of $F_{ST}$ because it does not require to define populations in advance and can even be applied in the absence of well-defined populations.

\begin{figure}[ht!]
	\centering
		\includegraphics[width=14cm]{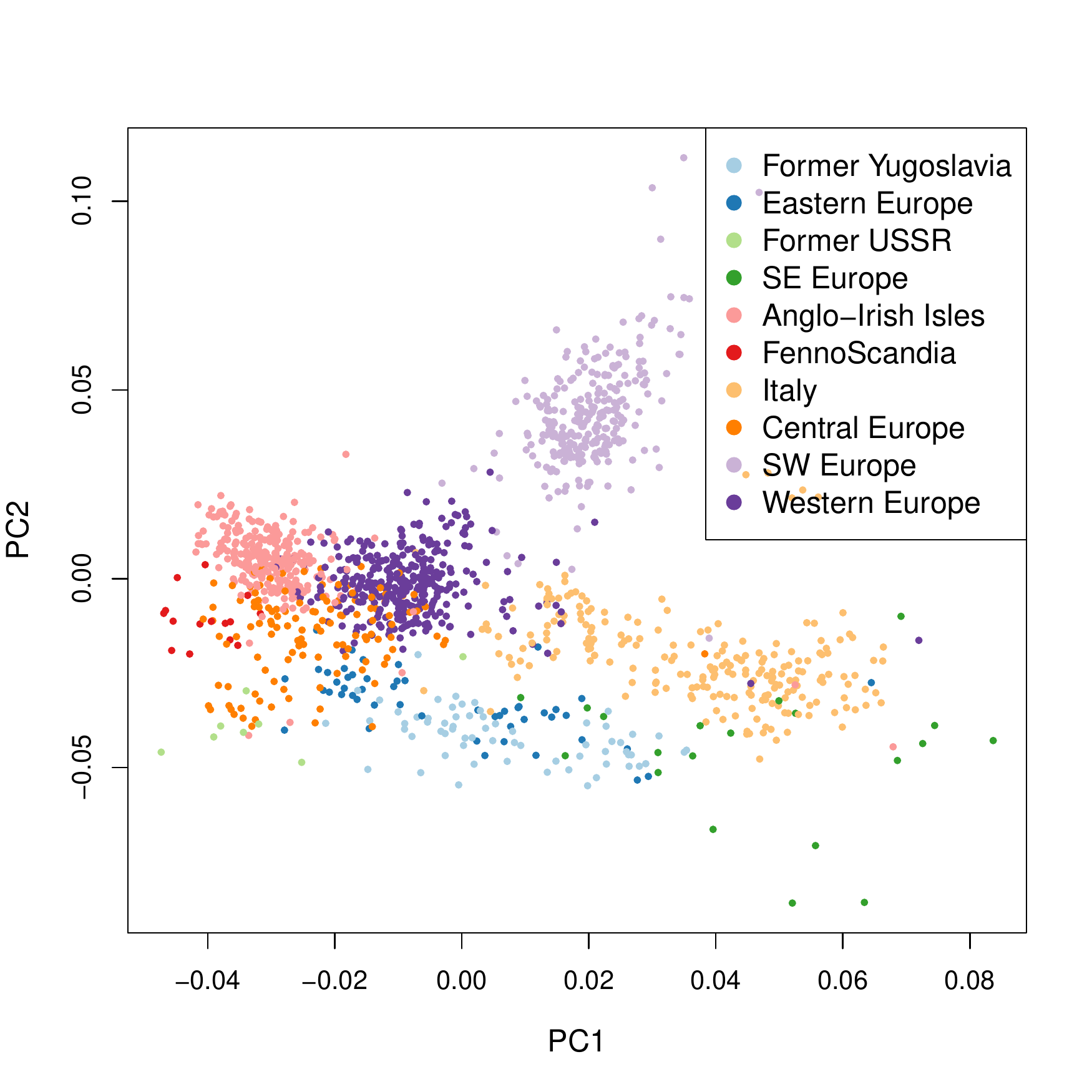}
	\caption{Principal component analysis with $K=2$ applied to the POPRES data.}
\label{fig:pca_popres}.
\end{figure}

\begin{figure}[ht!]
	\centering
		\includegraphics[width=14cm]{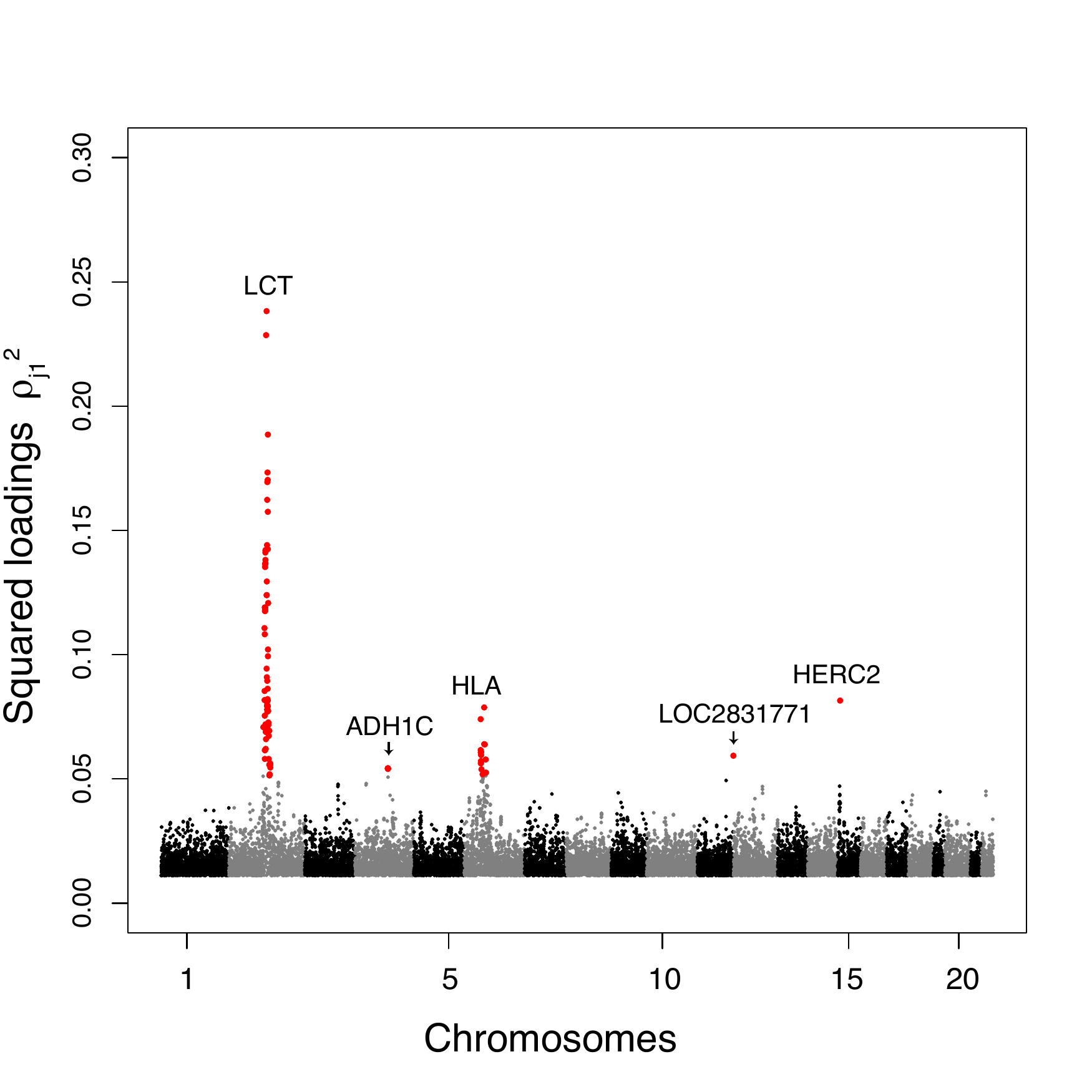}
	\caption{Manhattan plot for the POPRES data of the squared loadings $\rho_{j1}^2$ with the first principal component. For sake of presentation, only the top-ranked SNPs (top $5\%$) are displayed and the 100 top-ranked SNPs are colored in red.}
\label{fig:scan_popres}.
\end{figure}

To provide an example of genome scans based on PCA  when there are no clusters of populations, we additionally consider the POPRES data consisting of 447,245 SNPSs typed for 1,385 European individuals \cite[]{nelson08}. The scree plot indicates that there are $K=2$ relevant clusters (Fig. S3). The first principal component corresponds to a Southeast-Northwest gradient and the second one discriminates individuals from Southern Europe along a East-West gradient \cite[]{novembre08b,jay13} (Figure \ref{fig:pca_popres}). Considering the 100 SNPs most correlated with the first PC, we find that 75 SNPs are in the lactase region, 18 SNPs are in the HLA region, 5 SNPs are in the ADH1C gene, 1 SNP is in HERC2 and another is close to the LOC283177 gene (Figure \ref{fig:scan_popres}). When considering the 100 SNPs most correlated with the second PC,  we find less clustering than for PC1 with more peaks (Fig. S13). The regions that contain the largest number of SNPs in the top 100 SNPs are the HLA region (41 SNPs) and a region close to the NEK10 gene (10 SNPs), which is a gene potentially involved in breast cancer \cite[]{ahmed09}. The genome scan retrieves well-known signals of adaption in humans that are related to lactase persistence (LCT) \cite[]{bersaglieri04}, immunity (HLA), alcohol metabolism (ADH1C) \cite[]{han07} and pigmentation (HERC2) \cite[]{wilde14}. The analysis of the POPRES data shows that genome scan based on PCA can be applied when there is a clinal or continuous pattern of population structure without well-defined clusters of individuals.

When there are clusters of populations, we have shown with simulations that genome scans based on $F_{ST}$ can be reproduced with PCA. Genome scans based on PCA have the additional advantage that a particular axis of genetic variation, which is related to adaptation, can be pinpointed. Bearing some similarities with PCA, performing a spectral decomposition of the kinship matrix has been proposed to pinpoint populations where adaptation took place \cite[]{fariello13}.  However, despite of some advantages, the statistical problems related to genome scans with $F_{ST}$ remain. The drawbacks of $F_{ST}$ arise when there is hierarchical population structure or range expansion because  $F_{ST}$ does not account  for correlations of allele frequencies among subpopulations \cite[]{bierne13,lotterhos14}. An alternative presentation of the issues arising with $F_{ST}$ is that it implicitly assumes either a model of instantaneous divergence between populations or an island-model \cite[]{bonhomme10}. Deviations from these models severely impact false discovery rates \cite[]{duforet14}. Viewing $F_{ST}$ from the point of view of PCA provides a new explanation about why $F_{ST}$ does not provide an optimal ranking of SNPs for detecting selection. The statistic $F_{ST}$ or the proposed $h^2$ communality statistic are mostly influenced by the first principal component and the relative importance of the first PC increases with the difference between the first and second eigenvalues of the covariance matrix of the data. Because the first PC can represent ancient adaptive events, especially under population divergence models \cite[]{mcvean09}, it explains why $F_{ST}$ and the communality $h^2$ are  biased toward ancient evolutionary events. Following recent developments of $F_{ST}$-related statistics that account for hierarchical population structure \cite[]{bonhomme10,gunther13,foll14}, we proposed an alternative statistic $h^{\prime2}$, which should give equal weights to the different PCs. However, analyzing simulations and the 1000 Genomes data shows that $h^{\prime2}$ do not properly account for hierarchical population structure because outliers identified by $h^{\prime2}$ are almost always related to the last PC kept in the analysis. To avoid to bias data analysis in favor of one principal component, it is possible to perform a genome scan for each principal component.

In addition to ranking the SNPs when performing a genome scan, a threshold should be chosen to extract a list of outlier SNPs. We do not have addressed the question of how to choose the threshold and rather used empirical threshold such as the $99\%$ quantile of the distribution of the test statistic (top $1\%$). If interested in controlling the false discovery rate, we can assume that the loadings $\rho_{kj}$ are Gaussian with zero mean \cite[]{galinsky15}. Because of the constraints imposed on the loadings when performing PCA, the variance of the $\rho_{kj}$'s is equal to the proportion of variance explained by the $k^{\rm th}$ PC, which is given by $\lambda_k/(p\times (n-1))$ where $\lambda_k$ is the $k^{\rm th}$ eigenvalue of the matrix $Y Y^T$. Assuming a Gaussian distribution for the loadings, the communality (equation (\ref{eq:h})) can then be approximated by a weighted sum of chi-square distribution. Approximating a weighted sum of chi-square distribution with a  chi-square distribution, we have  \cite[]{yuan10}
\begin{equation}
\label{eq:chi2}
h^2 \times K/c \leadsto \chi^2_K,
\end{equation}
where $c=\sum_{i=1}^K \lambda_K/(p\times(n-1))$ is the proportion of variance explained by the first $K$ PCs. The chi-square approximation of equation (\ref{eq:chi2}) bears similarity with the approximation of \citet{lewontin73} that states that $F_{ST} \times ({\rm pop}-1)/ \bar{F}_{ST}$ follows a chi square approximation with $n$ degrees of freedom where $\bar{F}_{ST}$ is the mean $F_{ST}$ over loci and ${\rm pop}$ is the number of populations. In the simulations of an island model and of a divergence model, quantile-to-quantile plots indicate a good fit to the theoretical chi-square distribution of expression (\ref{eq:chi2}) (Figure S14). When using the chi-square approximation to compute P-values, we evaluate if FDR can be controlled using Benjamini-Hochberg correction \cite[]{benjamini95}. We find that the actual proportion of false discoveries corresponds to  the target FDR for the island model but the procedure is too conservative for the divergence model (Figure S15). For instance, when controlling FDR at a level of $25\%$, the actual proportion of false discoveries is of $15\%$. A recent test based on $F_{ST}$ and a chi-square approximation was also found to be conservative \cite[]{Lotterhos15}.

Analysing the phase 1 release of the 1000 Genomes data demonstrates the suitability of a genome scan based on PCA to detect signals of positive selection.  We search for variants extremely correlated with the first PC, which corresponds to differentiation between Africa and Eurasia and with the second PC, which corresponds to differentiation between Europe and Asia. For variants most correlated with the second PC, there is a significant enrichment of genic and non-synonymous SNPs whereas the enrichment is less detectable for variants related to the first PC. The enrichment analysis confirms that positive selection may favor local adaptation of human population by increasing differentiation in genic regions especially in non synonymous variants \cite[]{barreiro08}. Consistent with LD, we find that  candidate variants are clustered along the genome with a larger clustering for variants correlated with the Europe-Asia axis of differentiation (PC2). The difference of clustering illustrates that statistical methods based on LD for detecting selection will perform differently depending on the time frame under which adaptation had the opportunity to occur \cite[]{sabeti06}. The fact that population divergence, and its concomitant adaptive events, between Europe and Asia is more recent that the out-of-Africa event is a putative explanation of the difference of clustering between PC1 and PC2 outliers. Explaining the difference of enrichment between PC1 and PC2 outliers is more difficult. The weaker enrichment for PC1 outliers can be attributed either to a larger number of false discoveries or to a larger importance of other forms of natural selection such as background selection \cite[]{hernandez11}.

When looking  at the 100 SNPs most correlated with PC1 or PC2, we find genes for which selection in humans was already documented (9/24 for PC1 and 5/14 for PC2, Table S9). Known targets for selection include genes involved in pigmentation (MATP, OCA2  for PC1 and SLC45A2, SLC24A5, and MYO5C for PC2), in the regulation of sweating (EDAR for PC2), and in adaptation to pathogens (DARC, SLC39A4, and VAV2 for PC1).  A 100 kb region in the vicinity of the APPBPP2 gene contains one third of the 100 SNPs most correlated with PC1. This APPBPP2 region is a known candidate for selection and has been identified by looking for miRNA binding sites with extreme population differentiation \cite[]{li12}. APPBPP2 is a nervous system gene that has been associated with Alzheimer disease, and it may have experienced a selective sweep \cite[]{williamson07}. For some SNPs in APPBPP2, the differences of allele frequencies between Eurasiatic population and SubSaharan populations from Africa are of the order of $90\%$ (\url{http://www.popgen.uchicago.edu/ggv}) calling for a further functional analysis. Moreover, looking at the 100 SNPs most correlated with PC1 and PC2 confirms the importance of non-coding RNA (FAM230B, D21S2088E, LOC100133461, LINC00290, LINC01347, LINC00681), such as miRNA (MIR429), as a substrate for human adaptation \cite[]{li12,grossman13}.  Among the other regions with a large number of candidate SNPs, we also found the RTTN/CD226 regions, which contain many SNPs correlated with PC1. In  different selection scans, the RTTN genes has been detected \cite[]{carlson05,barreiro08}, and it is involved in the development of the human skeletal system \cite[]{wu10}. An other region with many SNPs correlated with PC1 contains the ATP1A1 gene involved in osmoregulation and associated with hypertension \cite[]{gurdasani15}. The regions containing the largest number of SNPs correlated with PC2 are well-documented instances of adaptation in humans and includes the EDAR, SLC24A5 and SLC45A2 genes. The KCNMA1 gene contains 7 SNPs correlated with PC2 and is involved in breast cancer and obesity \cite[]{oeggerli12,jiao11}.  As for KCNMA1, the MYO5C has already been reported in selection scans although no mechanism of biological adaption has been proposed yet \cite[]{chen10,fumagalli10}. To summarize, the list of most correlated SNPs with the PCs identifies well-known genes related to biological adaptation in humans (EDAR, SLC24A5,SLC45A2, DARC), but also provides candidate genes that deserve further studies such as the APPBPP2, TP1A1, RTTN, KCNMA1 and MYO5C genes, as well as the ncRNAs listed above.

We also show that a scan based on PCA can also be used to detect more subtle footprints of positive selection. We conduct an enrichment analysis that detects polygenic adaptation at the level of biological pathways  \cite[]{daub13}. We find that genes in the beta-defensin pathway are enriched in SNPs correlated with PC1. The beta-defensin genes are key components of the innate immune system and have evolved through positive selection in the catarrhine primate lineages \cite[]{hollox08}. As for the HLA complex, some beta-defensin genes (DEFB1, DEFB127) show evidence of long-term balancing selection with major haplotypic clades coexisting since millions of years \cite[]{cagliani08,hollox08}.  We also find that genes in the omega fatty acid oxidation pathways are enriched in SNPs correlated with PC2. This pathway was also found when investigating polygenic adaptation to altitude in humans \cite[]{foll14}. The proposed explanation was that omega oxidation  becomes a more important metabolic pathway when beta oxidation is defective, which can occur in case of hypoxia \cite[]{foll14}. However, this explanation is not valid in the context of the 1000 Genomes data when there are no populations living in hypoxic environments. Proposing phenotypes on which selection operates is complicated by the fact that the omega fatty acid oxidation pathway strongly overlaps with two other pathways: ethanol oxidation and glycolysis. Evidence of selection on the alcohol dehydrogenase locus have already been provided \cite[]{han07} with some authors proposing that a lower risk for alcoholism might have been beneficial after rice domestication in Asia \cite[]{peng10}. This hypothesis is speculative and we lack a confirmed biological mechanism explaining the enrichment of the fatty acid oxidation pathway. More generally, the enrichment of the beta-defensin and of the omega fatty acid oxidation pathways confirms the importance of pathogenic pressure and of metabolism in human adaptation to different environments \cite[]{hancock08,barreiro09,fumagalli11,daub13}.

In conclusion, we propose a new approach to scan genomes for local adaptation that works with individual genotype data. Because the method is efficiently implemented in the software {\it PCAdapt}, analyzing $36,536,154$ SNPs took only $502$ minutes using a single core of an Intel(R) Xeon(R) (E5-2650, 2.00GHz, 64 bits). Even with low-coverage sequence data (3x), PCA-based statistics retrieve well-known examples of biological adaptation which is encouraging for future whole-genome sequencing project, especially for non-model species, aiming at sampling many individuals with limited cost. 

\section*{Materials and Methods}
\subsection*{Simulations of an island model}
Simulations were performed with {\it ms} \cite[]{hudson02}. We assume that there are 3 islands with $100$ sampled individuals in each of them. There is a total of $1,400$ neutral SNPs, and $100$ adaptive SNPs. SNPs are assumed to be unlinked. To mimic adaptation, we consider that adaptive SNP have a migration rate smaller than the migration rate of neutral SNPs ($4N_0m=4$ for neutral SNPs) \cite[]{bazin10}. The strength of selection is equal to the ratio of the migration rates of neutral and adaptive SNPs. Adaptation is assumed to occur in one population only. The {\it ms} command lines for neutral and adaptive SNPs are given below (assuming an effective migration rate of $4 N_0 m = 0.1$ for adaptive SNPs).
\begin{verbatim}
./ms 300 1400 -s 1 -I 3 100 100 100 -ma x 4 4 4 x 4 4 4 x #neutral
./ms 300 100 -s 1 -I 3 100 100 100 -ma x 0.1 0.1 0.1 x 4 0.1 4 x #outlier
\end{verbatim}
The values of migrations rates we consider for adaptive SNPs are  $4 N_0 m = 0.04, 0.1, 0.4, 1, 2$.

\section*{Simulations of divergence models}
We assume that each population has a constant effective population size of $N_0= 1,000$ diploid individuals, with 50 individuals sampled in each population. The genotypes consist of 10,000 independent SNPs. The simulations were performed in two steps. In the first step, we used the software {\it ms} to simulate genetic diversity \cite[]{hudson02} in the ancestral population. We kept only variants with a minor allele frequency larger than $5\%$ at the end of the first step. The second step was performed with {\it SimuPOP} \cite[]{peng05} and simulations were started using  the allele frequencies generated with {\it ms} in the ancestral population. Looking forward in time, we consider that there are $100$ generations between the initial split and the following split between the two $B$ subpopulations, and $200$ generations following the split between the two $B$ subpopulations. We assume no migration between populations. In the simulation of Fig. \ref{fig:stacked_bar}, we assume that 250 SNPs confer a selective advantage in the branch leading to population $A$ and 250 other SNPs confer a selective advantage in the branch leading to population $B_1$. We consider an additive model for selection with a selection coefficient of $s=1.025$ for heterozygotes. For the simulation of Fig. \ref{fig:stacked_bar2}, we assume that there are four non-overlapping sets of $125$ adaptive SNPs with each set being related to adaptation in one of the four branches of the divergence tree. A SNP can confer a selective advantage in a single branch only. 

When including migration, we consider that there are $200$ generations between the initial split and the following split between the two $B$ subpopulations, and $100$ generations following the split between the two $B$ subpopulations. We consider migration rates ranging from $0.2\%$ to $5\%$ per generation. Migration is assumed to occur only after the split between $B_1$ and $B_2$. The migration rate is the same for the three pairs of populations. To estimate the $F_{ST}$ statistic, we consider the estimator of Weir and Cockerham \cite[]{weir84}.

\section*{1000 Genomes data}
We downloaded the 1000 Genomes data (phase 1 v3) at \url{ftp://ftp.1000genomes.ebi.ac.uk/vol1/ftp/phase1/analysis_results/integrated_call_sets/} \cite[]{Altshuler12}. We kept low-coverage genome data and excluded exomes and triome data to minimize variation in read depth. Filtering the data resulted in a total of $36,536,154$  SNPs that have been typed on $1,092$ individuals. Because the analysis focuses on biological adaptation that took place during the human diaspora out of Africa, we removed recently admixed populations (Mexican, Columbian, PortoRican, and AfroAmerican individuals from the Southwest of the USA). The resulting dataset contains 850 individuals coming from Asia (two Han Chinese and one Japanese populations), Africa (Yoruba and Luhya) and Europe (Finish, British in England and Scotland, Iberian, Toscan, and Utah residents with Northern and Western European ancestry). 

\section*{Enrichment analyses}

We used {\it Gowinda} \cite[]{kofler12} to test for  enrichment of Gene Ontology (GO). A gene is considered as a candidate if there is at least one of the most correlated SNPs (top $1\%$) that is mapped to the gene (within an interval of 50Kb upstream and downstream of the gene). Enrichment was computed as the proportion of genes containing at least one outlier SNPs among the genes of the given GO category that are present in the dataset. In order to sample a null distribution for enrichment, {\it Gowinda} performs resampling without replacement of the SNPs. We used the {\it --gene} option of {\it Gowinda} that assumes complete linkage within genes. 

We performed a second enrichment analysis to determine if outlier SNPs are enriched for genic regions. We computed odds ratio \cite[]{kudaravalli09}
$$
{\rm OR}= \frac{{\rm Pr}({\rm genic} | {\rm outlier})}{{\rm Pr}({\rm not \, genic} | {\rm outlier})}  \frac{{\rm Pr}({\rm not \, genic} | {\rm not \, outlier})}{{\rm Pr}({\rm genic} | {\rm not \, outlier})}. 
$$
We implemented a permutation procedure to test if an odds ratio is significantly larger than 1 \cite[]{fagny14}. The same procedure was applied when testing for enrichment of UTR regions and of non-synonymous SNPs.

\section*{Polygenic adaptation}

To test for polygenic adaptation, we determined whether genes in a given biological pathway show a shift in the distribution of the loadings \cite[]{daub13}. We computed the SUMSTAT statistic for testing if there is an excess of selection signal in each pathway \cite[]{daub13}.  We applied the same pruning method to take into account redundancy of genes within pathways. The test statistic is the squared loading standardized into a z-score \cite[]{daub13}. SUMSTAT is computed for each gene as the sum of test statistic of each SNP belonging to the gene. Intergenic SNPs are assigned to a gene provided they are situated 50kb up or downstream. We downloaded 63,693 known genes from the UCSC website and we mapped SNPs to a gene if a SNP is  located within a gene transcript or within 50kb of a gene. A total of 18,267 genes were mapped with this approach. We downloaded 2,681 gene sets from the NCBI Biosystems database. After discarding genes that were not part of the aforementioned gene list, removing gene sets with less than 10 genes and pooling nearly identical gene sets, we kept 1,532 sets for which we test if there was a shift of the distribution of loadings. 

\subsection*{Acknowledgments} This work has been supported by the LabEx PERSYVAL-Lab (ANR-11-LABX-0025-01) and the ANR AGRHUM project (ANR-14-CE02-0003-01). POPRES data were obtained from dbGaP (accession number phs000145.v1.p1)

\bibliographystyle{mbe}
\renewcommand\refname{References}
\bibliography{fastpcadapt_bib}

\clearpage

\clearpage

\clearpage


\clearpage


\clearpage


\clearpage

\begin{table}[h]
  \begin{tabular}{c | c c c c}
  ~  &    	${\rm top}\, 0.1\%$ & $ {\rm top}\, 0.01\%$ & ${\rm top}\, 0.005\%$ &top 100 SNPs\\
  \hline
pc1 - genic/nogenic	& $1,60^{*}$ & 1,24 & 1,09 & 1,93\\
pc1 - nonsyn/all	 & 1,70 & 1,18	& 2,42 & $10,07^{*}$\\
pc1 - UTR/all & 1,37	& 0,80 & 1,65 & 3,44\\
  \hline
pc2 - genic/nogenic	& $1,51{*}$ & 2,27 & $4,73 ^{**}$ & $4,44^{*}$\\
pc2 - nonsyn/all	 & 1,72 & $4,66^{*}$ & 7,40 & $12,18^{*}$\\
pc2 - UTR/all & 1,68 & $4,01^{*}$ & 3,36 & 2,73
  \end{tabular}
  \caption{Enrichment measured with Odds Ratio (OR) of the variants most correlated with the principal components obtained from the 1000 Genomes data. Enrichment significant at the $1\%$ (resp. $5\%$) level are indicated with $^{**}$ (resp. $^{*}$).}
\label{tab:1}
\end{table}

\end{document}